**4D-Printing Assisted Scaffolds to Form Cardiac Bricks[1]**

Hossein Goodarzi Hosseinabadi [1,2*]

[1] Institute of Pharmacology and Toxicology, University Medical Center Göttingen, Robert-Koch-Str. 40, 37075 Göttingen, Germany.
[2] Department of Biomedical Engineering, Technical University of Eindhoven, STO 4.37, Bio-medical Engineering (BmE) department, Eindhoven, 5612 AZ, Netherlands (present address)

Correspondance: HGH, **h.goodarzi.hosseinabadi@tue.nl**

**Abstract**
Myocardial infarction causes myocardium thinning, fibrosis, and progressive heart failure. Engineered human myocardium (EHM) is tested clinically as a first-in-class product for sustainable remuscularization in patients with advanced heart failure. Current EHM production procedure from iPSC-derived cardiomyocytes and stromal cells, is time consuming and involves thin constructs. Here, I introduce 4D-DLP-printed foldable scaffolds with potential to create modular cylindrical cardiac bricks. This enables self-assembly into thicker and aligned sarcomeres with synchronous contractility mimicking a native myocardium. When optimized and integrated with cryopreservation protocols, the biomanufacturing and biobanking of these cellular building blocks may overcome current EHM limitations and advance translational regenerative therapies for myocardial infarction. The structure-material properties investigations into these new class of life building blocks paves the way for future medical breakthrough.

**Introduction**
Cardiovascular disease accounts for 32% of global deaths and is primarily caused by ischemic heart disease and myocardial infarction. Reduction of mortality after myocardial infarction has paradoxically led to an increase in heart failure. Myocardial infarction causes partial loss of the heart cells and progressive scar formation leading to thinner myocardium till heart failure. The heart's limited capacity to regenerate means patients typically progress from an asymptomatic state to clinically overt heart failure with a 50% mortality rate only in five years [1]. Given these daunting numbers and the lack of treatment options, there is unmet medical need to develop new strategies for regenerative and reparative therapies, such as engineered human myocardium (EHM) [2]. EHM technology is presently tested clinically as a first-in-class product from induced pluripotent stem cells (iPSCs) for sustainable remuscularization in patients with advanced heart failure [3,4].
The irreversible cardiomyocyte loss and scar formation due to myocardial infarction can be treated with transplantation of the EHM. The state-of-the-art thin layer of EHM is prepared from iPSC-derived cardiomyocytes (CMs) and stromal cells (StCs) [2]. CMs cause tissue contraction thus they develop the contractile forces, while StCs cause tissue compaction beside secreting extracellular matrix proteins such as collagens to support 3D organization of EHM and to bind extracellular matrix via integrins. However, conventional EHMs are composed of thin layers of collagen-based tissue with a random orientation and cellular architecture (**Fig. 1a**), hardly replicating sufficient thickness, anisotropic architecture, and vascularization needed for functionally mature myocardium.

---

[1] *ArXiv preprint version, prepared for Scr. Mater.*

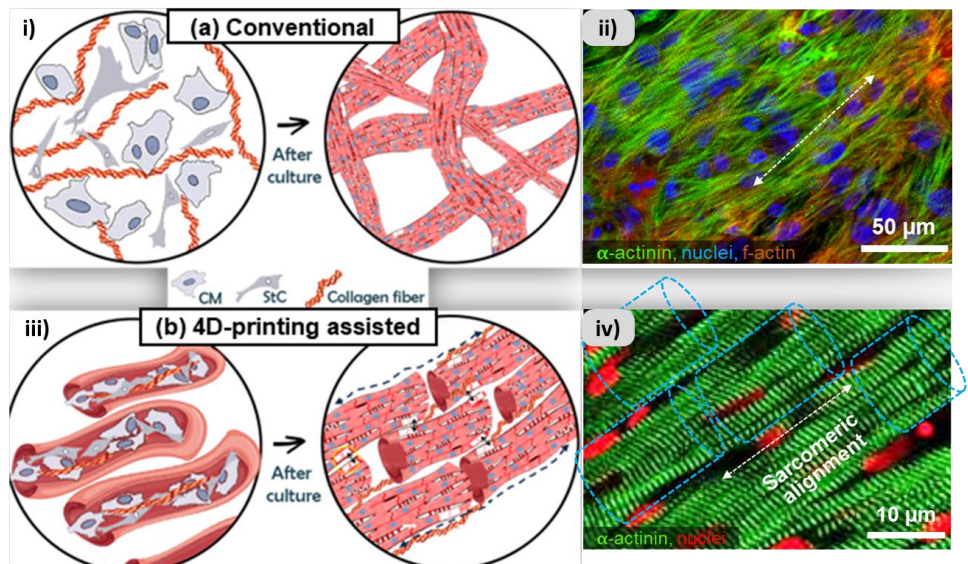

**Figure 1. Strategies to create myocardium tissue**: (i) Conventional EHM production scheme (a); (ii) Sarcomere alignment in small-size EHM ring produced from iPSC-derived CM:Fibroblast (70%:30%) after 7 days of culture in the lab using the conventional EHM production protocol (adopted from [2]); (iii) The proposed 4D-DLP-printing assisted EHM production scheme to allow generation of thicker myocardium using pre-formulated cellular cylinders (cardio-bundles) as structural units or cellular-bricks; (iv) α-actinin sarcomere alignment in a natural myocardium of a guinea pig (adopted from [5]) showing more uniform distribution of cells at the cross-section of the myocardium tissue; hashed blue lines are to resemble that the native architecture is obtainable by self-assembly of the proposed cylindrical units.

Being time consuming, expensive, and the lack of alignment control can limit the efficacy and scalability of current EHM production process. To address these challenges, here a strategy is developed based on using 4D-DLP-printed hydrogels as light-curable foldable material [6] to form EHM bricks with a cylindrical shape. These biocompatible foldable scaffolds serve as a shape reconfigurable tool to create cardiac building blocks (i.e. cardiac bricks) that can be produced in scale, cryo-preserved, and be self-assembled into thicker aligned myocardium tissues (**Fig. 1b-iii**) which closely replicate the natural myocardium cellular architecture (**Fig. 1b-iv**). The conceptual framework of the proposed approach and its comparison to native myocardium are illustrated in **Fig. 1**. As illustrated, here the focus is placed on a strategy enabling scalable fabrication of cellular bricks to enable potential self-organization into thicker EHMs. This modular strategy is based on using cylindrical scaffolds fabricated by digital light processing (DLP) of a foldable polyurethane-based photopolymer (PUPEGDA), which undergoes 4D shape transformation into tubular geometries. These cardiac bricks serve as building blocks with potential to self-assemble into thicker and anisotropic myocardium. Compared to scaffold-free EHM, the approach shall enable precise control of geometry and porosity as well as possibility of ultrasound-mediated alignment of multiple bricks into ordered assemblies. The novelty of this method is in combining 4D-printing, self-folding polymer films, and human iPSC-derived cardiomyocytes to generate anisotropic building blocks that can be assembled into functionally relevant tissue, surpassing the performance of existing scaffold creation strategies.

**Fig. 2** shows the DLP printing setup and the workflow for fabrication of scaffolds with grooved, porous, and foldable designs.

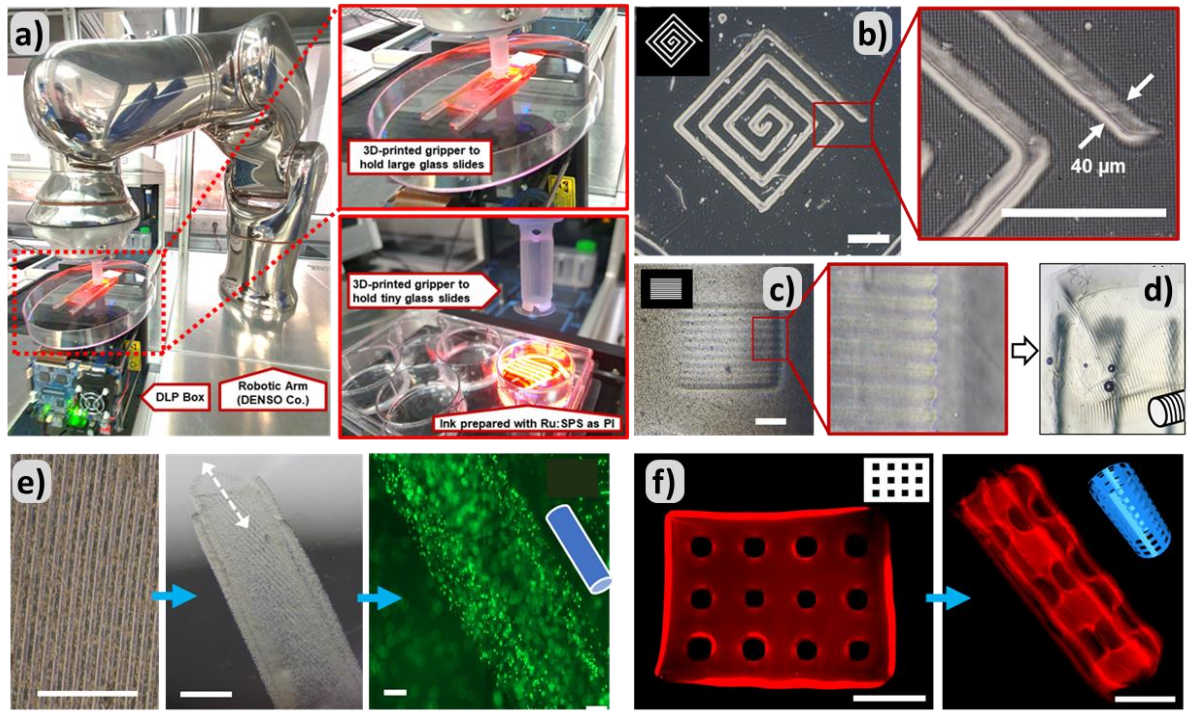

**Figure 2.** DLP printing setup used for 4D-printing of the scaffolds: **a)** A DLP box is integrated with a robotic arm under a sterile hood for fabrication of patterned hydrogels at high-resolution according to good manufacturing practice (GMP) principles: **b)** High resolution channels are created using the photo-crosslinkable ink showing both the optical hardware and the refractive-index of the ink perfectly match the requirements for high resolution DLP printing (each small rectangle at enlarged image shows one DLP voxel of 7 µm length; inset: 200 µm); **c)** A rectangular hydrogel created with grooves on the surface (hills and valleys guide the cellular organization); the patterned films can self-fold to create a permeable cylindrical hydrogel scaffold, **d)** with grooves perpendicular to the cylinder axis **e)** with grooves parallel to the cylinder axis (the ink DLP printed to create a flat film with groove and ridge pattern, then coated and seeded with CM- HFF master mix and coumarin 6, then self-folded to create a 3D tubular scaffold, revealing cellular activity at day 7 post-print), **f)** or with programmed mesh of 3×4 micro-porosities (adopted from [6]) after immersion in culture media (insets: 500 µm).

To fabricate the self-foldable scaffold constructs, PUPEGDA was synthesized as previously described [6] and printed with a custom-built DLP setup. In brief, to prepare PUPEGDA a two-step polymerization started with mixing polyethylene-glycol (Mn: 3000 Da) and hexamethylene diisocyanate for copolymerization reaction, which subsequently terminated by adding hydroxyethyl acrylate (HEA) and a catalyst at 80°C under inert atmosphere. The solution was precipitated in 100% IPA, centrifuged, dialyzed for one week, freeze-dried, and stored in dark and cold conditions. LAP (0.1 w/v%) [7] and Ru:SPS (3:30 mM) [6] were used as photo-initiators in aqueous solution and a commercially available DLP box (1050 × 920 pixels, EKB Technologies™) was customized with UV optical lenses and integrated with a robotic arm

(DYNSO™), installed under a sterile hood to enable automated scaffold fabrication with precise XYZ control (**Fig. 2a**). The system achieved a theoretical lateral printing resolution of 7 µm, defined by the DLP mirror pitch. Photocrosslinking was performed within a fluidic chip under ~20 mW/cm² light intensity, enabling reproducible fabrication of scaffolds with an operational lateral resolution over 40 µm (**Fig. 2b**).

Incorporating grooves (**Fig. 2c,d,e**) and porous features (**Fig. 2f**) into these scaffolds has potential to enhance cell alignment and mass transport, respectively, to increase cell viability, waste removal, and oxygenation. These EHM bricks have potential to be further directed and aligned using ultrasound-mediated cues within a collagen matrix for imposing in vivo in a clinical setting to regulate the native-like myocardium alignment in heart infarction patients. This approach offers novelty compared to other methods such as pull-spinning, rotary jet spinning, and extrusion bioprinting [8] through enabling modular, anisotropic self-assembly into thick EHM with improved functional properties. For instance, pull-spinning [7] and focused rotary jet spinning [8] methods are already used to create anisotropic cardiac tissue scaffolds. The limited surface area of these scaffolds and incompatibility of thicker scaffolds with myocardium mechano-physiology has limited application of these scaffold-based methods to scale thick tissue production. These studies, however, revealed possibility of controlling the magnitude and direction of contractile forces in cardiac tissues by scaffold engineering. The cellular orientation in many tissues, including cardiac tissues, determines function or dysfunction [8]. Helical cells in heart morphology are speculated to be deterministic in reaching physiological pumping efficiencies [8]. Different cardiac scaffolds are developed and integrated with disease-relevant and patient-specific cells, such as differentiated iPSCs, to leverage the utility of the EHM as a screening tool [9]. Macqueen et al. [7] introduced a tissue-engineered model of the heart ventricle based on the pull-spinning of scaffolds. Chang et al. [8] developed focused rotary jet spinning technique to enable continuous deposition of polymeric thin fibers with a sort of preferred orientation at the micrometer scale. The scaffolds were fabricated with different alignments and seeded with contractile cells. To validate how cell misorientation affects heart dysfunction, helical fibers were fabricated which led to replicating normal heart function, and deliberately misaligned fibers were fabricated leading to reveal diseased heart function. Williams et al. [10] reported fabrication of a large scale conical human ventricular model by seeding the cells on a groove-shape scaffold. First, a patterned scaffold sheet was made of a photocrosslinkable polyurethane using a master-mold composed of grooves. CMs were mixed in a fibrin solution and seeded two consecutive times into the flexible grooved sheets of scaffold, and then placed into a conical mold for maturation. Generation of tissue engineered ventricles was reported as well as the observation that pre-aligned CMs outperformed those randomly oriented structures. Two decades ago, Sefton and colleagues made attempts to create sub-millimeter collagen structures containing embedded cells for modular assembly to create tissues such as pancreatic islets [11], muscle micro-tissues [12], and perusable vascularized organoids [13]. Comparing the tissue displacement and synchronicity in these studies reveals that well aligned cardiac tissues can generate 4–10 times the contractile force of those with randomly oriented sarcomeres in miniaturized samples.

4D printing is an emerging scaffold fabrication technique [16,17] deviated from conventional 3D printing. Considering time as a fourth dimension (i.e. in 4D printing) allows the 3D printed structure to change the shape when stimulated with, for example, moisture [6], and give advantages compared to traditional 3D printing. For example, the conventional 3D printing of hollow tubular scaffolds with cells (e.g. inkjet, SLA, or DLP-based printing) mostly suffers from aggregation and non-uniform attachment of the cells after seeding the cells into the 3D tubes due to cell sedimentation [18] or lack of resolution due to light scattering from the cells membrane [16,19,20]. In 4D printing, after cell attachment to the patterned planar scaffolds the shape transformation permits the formation of tubular structures with evenly distributed cells on the inner wall of the tubular scaffold [21–23]. 4D DLP bioprinting in a fluidic chip can provide a superior speed for the fabrication of highly complex micro-architectures at high throughput and scalability [24]. One of the barriers preventing the development of 4D DLP bioprinting in cardiac tissue engineering is the limited access to proper set of inks not only compatible with iPSC-CMs but also capable of showing self-foldability post printing. Here, the toxicity of the purified PUPEGDA hydrogels (discussed in [6]) is examined. **Fig. 3a** demonstrates the formation of biocompatible hydrogel scaffolds without evident toxicity for the used iPSC-CM and HFF cell lines.

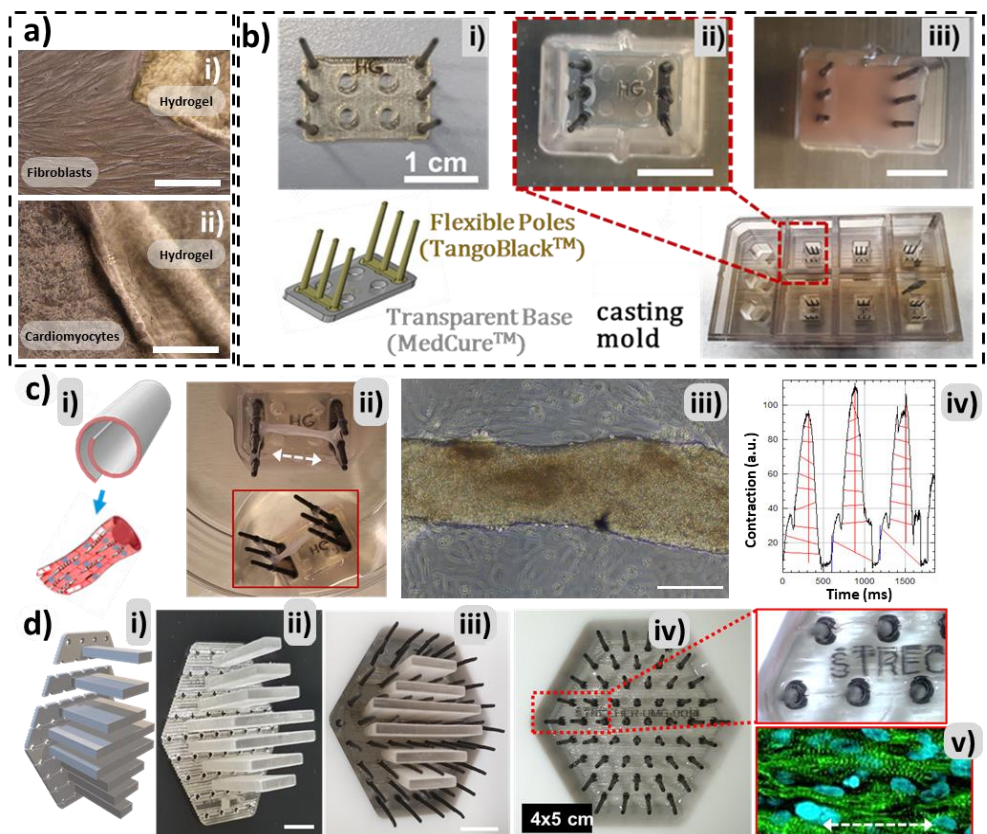

**Figure 3.** Cylindrical cardiac bricks exhibit anisotropic sarcomere alignment, almost uniform cell density, and substantial contractility. **a)** toxicity test for the DLP printed PUPEGDA films for iPSC-CM and HFF cells after 3 days of culture; **b)** Illustration of the CAD and the mold 3D-printed with PolyJet multi-material technique for casting the EHM tissues; **i)** the 3D-printed stretcher involves flexible poles composed of TangoBlack™ adopted from [2]; next, the standard EHM production procedure is followed by **ii)** covering the stretcher base with agarose then **iii)** CM-HFF-Collagen mix was casted; **c)** the tissue layer was exposed to the DLP-printed self-foldable PUPEGDA hydrogel film **(i)** to create a collagen-based cylindrical EHM; **(ii)** the perspective view of the formed EHM (day 1); **(iii)** photo of the formed EHM after 27 days of culture; **(iv)** the self-contraction profile of the illustrated cylindrical EHM; **d)** a modular (micro-) patterned stamp is **i)** CAD designed and **ii)** 3D printed with PolyJet technique (inset: 1cm), **iii,iv)** to pattern the agarose substrate on stretcher, within the Teflon mold previously used for clinical scale EHM casting (see ref. [2,4]) using **ACTN2-citrine** CMs (70%) and HFF (30%) cells in collagen; **v)** Confocal images captured at day 21 showed sarcomere alignment in the grooves direction (two-side arrow: sarcomere alignment).

Here, human induced pluripotent stem cell derived cardiomyocytes and stromal cells were prepared in a 70:30 ratio, following protocols adapted from Tiburcy et al. [2], and suspended in collagen type I prior to seeding. ROCK inhibitor (1:2000) was included when seeding to improve viability. After embedding the forming tissue in the foldable scaffold, the constructs were cultured with the standard maturation media [2] over 3 weeks aiming to observe if the EHM brick preserves a homogeneous cell distribution and sufficient oxygenation throughout the construct. **Fig. 3b** illustrates the adopted procedure used for conventional casting the EHM tissues. **Fig. 3c** shows the implementation of the foldable hydrogel scaffold to form the cylindrical cardiac bricks with aligned sarcomeres and uniform cell density, along with the representative tissue contractility profile measured by MUSCLEMOTION plug-in using ImageJ software on the video recorded by Zeis microscope. A supplementary video file illustrates the spontaneous and synchronous self-contraction and beating of a formed cylindrical EHM block (**Movie S1**). Confocal microscopy (**Fig. 3d-v**) demonstrated evidence of guided sarcomere organization, including alignment of ACTN2-citrine labeled cardiomyocytes along the micro-grooves direction on clinical size large EHM production (per discussed in [4]) (**Fig. 3d**).

Overall, results suggest that the use of 4D-printed scaffolds possibly accelerates the establishment of anisotropy and promotes the cardiomyocyte maturation potential. The incorporation of porosity and groove-guided alignment can enhances oxygenation and creates a microenvironment favorable for vascular ingrowth, potentially overcoming one of the major bottlenecks in thick tissue engineering. When assembled into a centimeter large patch with embedding grooves on the substrate (**Fig. 3d-iv**), the cylindrically thick cardiac brick maintains uniform cell density and contractility comparable to those reported for conventional EHM [2]. Moreover, employing thousands of these cylindrical cellular bricks with diameters in the hundreds of micrometers range is expected to exhibit greater potential for ultrasound-guided spatial alignment while maintaining inter-brick contact, compared to individual cardiomyocytes having only a few micrometers and remain separated by hundreds of micrometers under ultrasound alignment. This strategy introduces an additional biophysical guidance cue that may further enhance anisotropic tissue architecture during the fabrication of thicker engineered heart muscle (EHM) constructs.

Potential limitations of this strategy include restricted oxygen and nutrient diffusion in thicker constructs, difficulties in automated handling of large numbers of cylindrical EHMs, variable adhesion among different EHM bricks, and risks associated with the EHM damage due to incomplete scaffold separation. While porosity-engineered scaffolds may improve mass transport and tissue compatibility, further refinement of fabrication protocols remains necessary. Future work will focus on enhancing scaffold reproducibility and controlled degradability, alongside implementing dynamic culture systems, to advance the development of clinically translatable, scaffold-free engineered heart muscle.

In conclusion, cylindrical scaffolds produced by 4D-printing have the potential to provide modular cardiac bricks that can possibly be assembled into thick, anisotropic engineered human myocardium. These constructs can achieve improved sarcomere alignment, contractility, contractile forces, and conduction velocity values approaching those of native myocardium, thereby addressing key limitations of current engineered heart muscle technology. Despite current limitations for low reproducibility due to not yet optimized protocols, this strategy offers a promising path toward clinically relevant myocardial regeneration. Downscaling the method to single cell scaffolding by using two photon polymerization method [25] with improved resolutions in one or two order(s) of magnitude higher resolution in hundreds of nanometer range is another direction for the next research. Therefore, more systematic studies are needed to reveal structure-material properties investigations into this new method of life building block generation.

**Movie S1:** Supplementary video demonstrating contraction dynamics of a single cylindrical cardiac brick.

## Acknowledgements
HGH acknowledges the partial support from Alexander von Humboldt fellowship during performing this research as well as the supports from Prof. W.H. Zimmermann's group in the Institute of Pharmacology and Toxicology, University Medical Center Göttingen and the access to the iPSC facility and the ACTN2-citrine labeled cardiomyocytes in this research.